\documentclass[aps,prl,twocolumn,epsfig,showpacs,superscriptaddress,amsmath,amssymb,floatfix]{revtex4}
\usepackage{amssymb}
\usepackage{amsmath}
\usepackage{latexsym}
\usepackage{graphicx}
\usepackage{dcolumn}
\usepackage{bm}
\usepackage{natbib}
\DeclareGraphicsExtensions{.eps}

\newcommand{\be}{\begin{equation}}
\newcommand{\ee}{\end{equation}}
\newcommand{\ba}{\begin{eqnarray}}
\newcommand{\ea}{\end{eqnarray}}
\newcommand{\bea}{\begin{eqnarray*}}
\newcommand{\eea}{\end{eqnarray*}}

\begin{document}

\title{Nonlinear Sigma Model Analysis of the AFM Phase Transition of the Kondo Lattice}
\author{T.~Tzen Ong}
\affiliation{Department of Applied Physics, Stanford University, Stanford CA 94305, USA}
\affiliation{IBM Research Division, Almaden Research Center, San Jose, CA 95120, USA}
\author{B.~A.~Jones}
\affiliation{IBM Research Division, Almaden Research Center, San Jose, CA 95120, USA}
\date{\today}

\begin{abstract}
We have studied the antiferromagnetic quantum phase transition of a 2D Kondo-Heisenberg square lattice using the non-linear sigma model. A renormalization group analysis of the competing Kondo -- RKKY interaction was carried out to 1-loop order in the $\epsilon$ expansion, and a new quantum critical point is found, dominated by Kondo fluctuations. In addition, the spin-wave velocity scales logarithmically near the new QCP, i.e breakdown of hydrodynamic behavior. The results allow us to propose a new phase diagram near the AFM fixed point of this 2D Kondo lattice model.
\end{abstract}
\maketitle

The physics of heavy-fermion metals, systems that contain both localized $f$-electrons and conduction electron bands, has been of great interest in the strongly-correlated community for several decades. Initial work by Doniach \cite{DoniachPhysB77} pointed out the competition between the Kondo ($J_K$) and the RKKY ($J_H$) interaction, and the quantum critical point (QCP) associated with the transition from an antiferromagnetic metal (AFM) to a paramagnetic (PM) state as $J_K$ and $J_H$ vary. Work on the two-impurity model \cite{JonesPRL87, JonesPRL88} showed that the RKKY and Kondo coupling are not mutually exclusive either for ferromagnetic or antiferromagnetic interactions. In fact, a correlated Kondo effect is the rule rather than the exception, with the impurities partially compensated by the conduction electrons, and partially by each other (for AF interactions). In 1-D, there have been various numerical \cite{TsunetsuguRMP97} and analytical \cite{TsvelikPRL94, KivelsonPRL96} studies. However, the physics of higher dimensional systems and the corresponding quantum phase transition (QPT) is much less well-understood.

A commonly assumed scenario is that the $f$-electrons delocalize and are included in the Fermi surface, resulting in a large Fermi surface as given by Luttinger's Theorem \cite{LuttingerPR60, OshikawaPRL00}, and the QPT is of the spin density-wave (SDW) type. This QPT has typically been understood in the Hertz-Millis approach \cite{HertzPRB76, MillisPRB93}, and the physics of delocalized $f$-electrons forming a heavy fermi liquid have been well-understood in a large-N approach \citep{ReadJPhysC83, AuerbachPRL86, MillisPRB87}. However, experimental studies have shown that the Hertz-Millis picture may be inadequate for describing the behavior of the system near the QCP \cite{PaschenNature04, SchroderNature00}. An alternative picture of the AFM to PM transition is that the $f$-electrons do not delocalize and become part of the Fermi surface, hence the Fermi surface is small. This picture has been proposed by several groups \cite{SiIJMPB99, ColemanJPC01}, and studied within the DMFT approach \cite{SiIJMPB99,GrempelPRL03}. Exotic ground states have also been proposed for Kondo lattice systems \cite{AndreiPRL89, SenthilPRL03, LeePRB07} that include fractionalized quasi-particles.

\begin{figure}[bhp]
\begin{center}
\includegraphics[angle=0, width=4.5cm]{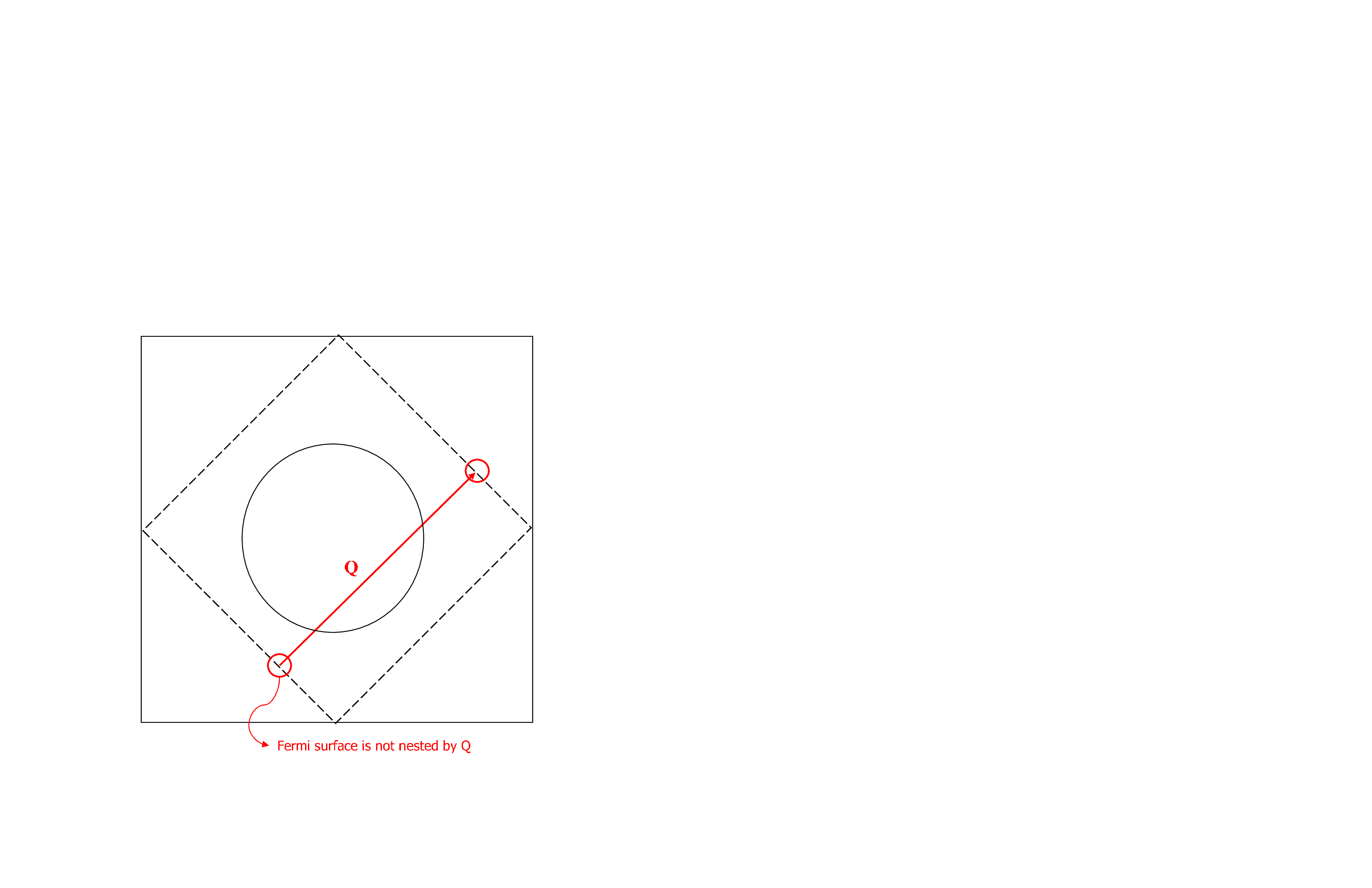}
\end{center}
\caption{Diagram of Fermi surface with no ``hot"-spots nested by $Q = (\pi,\pi)$. The solid black lines indicate the Fermi surface, and the dashed black lines indicate the magnetic B.Z. .}
\label{fig:FermiSurf}
\end{figure}

We study the effect of the Kondo coupling on the antiferromagnetic phase transition, and also see if the various exotic states can be obtained from a more microscopic approach. We assume that the AFM state has a small Fermi surface, with no``hot-spots" spanned by the Neel vector $(\pi,\pi)$, as shown in Fig. \ref{fig:FermiSurf}.  
To study the AFM QPT of the two-dimensional (2D) square Kondo lattice with a small Fermi surface, we map the system onto a non-linear sigma model (NLSM) coupled to conduction electrons. We then obtain an effective Lagrangian by integrating out the fermionic degrees of freedom, and carry out a renormalization group (RG) analysis to one-loop within the $\epsilon$-expansion. The calculations and results are described in the following sections.

As we shall show, there is a new QCP, with logarithmic scaling due to the Kondo interaction that plays an essential role in the QPT. The Kondo interaction strongly affects the spin-wave fluctuations near the QCP, and leads to logarithmic scaling of the spin-wave velocity, implying localization of the spin-waves due to the Kondo effect. Based upon the RG results, we also propose a possible phase diagram near the AFM state.

\begin{figure}[hbtp!]
\parbox{2cm}{
\includegraphics[angle=0, width=2cm]{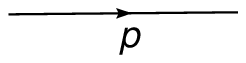}} 
\parbox{2cm}{
\begin{displaymath}
: \frac{c}{\omega^2 + c^2 |\vec{p}|^2}
\end{displaymath}
} \\
\parbox{2cm}{
\includegraphics[angle=0, width=2cm]{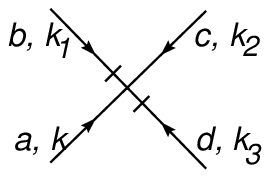}} 
\parbox{2cm}{
\begin{eqnarray*}
& : & -\frac{g}{c}(\omega_1 \omega_3 + c^2 \vec{k}_1 \cdot \vec{k}_3) \nonumber \\
 & & \times \delta^{ab} \delta^{cd} \nonumber \\
\end{eqnarray*}
} \\
\parbox{2.5cm}{
\includegraphics[angle=0, width=2.5cm]{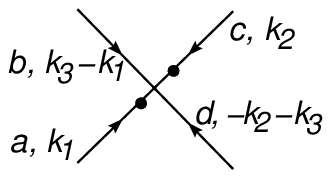}} 
\parbox{2.5cm}{
\begin{eqnarray*}
& : & - g_k \, \frac{|\omega_{k_3}|}{v_F |\vec{k}_3|} \left( \omega_1 \omega_2 \right) \nonumber \\
& & \times \epsilon^{eab} \epsilon^{ecd} \nonumber
\end{eqnarray*}
}
\caption{Feynman rules for the spin-wave propagator and two vertices. The third diagram is the vertex due to the Kondo interaction.}
\label{fig:FeynRules}
\end{figure}

The Hamiltonian for the Kondo-Heisenberg model is written as:
\begin{equation}
H = \sum_{\vec{k}} \epsilon_{\vec{k}} c_{\vec{k}}^{\dagger}c_{\vec{k}} + J \sum_{<i,j>} \vec{S}_{i} \cdot \vec{S}_j + K \sum_i \vec{S}_i \cdot \vec{S}_c(\vec{r}_i)
\end{equation}

where $J, K > 0$, are the Heisenberg and Kondo coupling respectively, and $\vec{S}_c(\vec{r}_i) = c_{i,\alpha}^{\dagger} \sigma_{\alpha \beta} c_{i,\alpha}$. Since we are interested in the AFM phase, the Heisenberg term can be mapped onto the non-linear sigma model in a well-known manner \cite{HaldanePRL88}. 
The local spin, $\vec{S}_i = (-1)^i \vec{n} + \frac{1}{|\vec{S}|\vec{l}}$, has both a Neel component, $\vec{n}$, and a small ferromagnetic moment, $\vec{l} \propto \partial_{\tau} \vec{n} \times \vec{n}$. The conduction electrons couple to both the Neel vector, $\vec{S}_c(\vec{r}_i) \cdot \vec{n}$, and also the ferromagnetic part of the local spin, $\vec{S}_c(\vec{r}_i) \cdot \vec{l} $. Upon integrating out the electrons, we obtain two 4th order terms, one due to the coupling to the Neel component and the other to the ferromagnetic part. As we have assumed a Fermi surface that is not nested by the AF wave vector, $(\pi, \pi)$, there is a kinetic energy gap for the former term. Similarly, there are no low-lying fermionic excitations that couple to $\vec{n}$ to give rise to Landau damping; hence it remains a $z = 1$ theory. The effective Lagrangian in terms of the Goldstone modes, $\pi^a$, where $a \in {x,y}$ is,
\vspace{-1.75cm}
\begin{widetext}
\begin{eqnarray}
L & = & \frac{1}{c} \int \frac{d \omega_1}{2 \pi} \frac{d^d \vec{k}_1}{(2 \pi)^d} (\omega_1^2 + c^2 |\vec{k}_1|^2) \pi^i(k_1) \pi^i(-k_1) \nonumber \\
  & & + \frac{g}{c} \int \frac{d \omega_1}{2 \pi}..\frac{d \omega_4}{2 \pi} \frac{d^d \vec{k}_1}{(2 \pi)^d}..\frac{d^d \vec{k}_4}{(2 \pi)^d} (i^2)(\omega_2 \, \omega_4 + c^2 \vec{k}_2 \cdot \vec{k}_4) \pi^a(k_1) \pi^a(k_2) \pi^b(k_3) \pi^b(k_4) \delta(k_1 + k_2 + k_3 + k_4) \nonumber \\     
   & & + g_k \int \frac{d \omega_1}{2 \pi}..\frac{d \omega_3}{2 \pi} \frac{d^d \vec{k}_1}{(2 \pi)^d}..\frac{d^d \vec{k}_3}{(2 \pi)^d} \frac{|\omega_3|}{v_F |\vec{k}_3|} \epsilon^{abe} \epsilon^{cde} (i \omega_1 \, i \omega_2) \pi^a(k_1) \pi^b(k_3 - k_1) \pi^c(k_2) \pi^d(-k_3 - k_2)
\end{eqnarray}
\end{widetext}
\vspace{-1.5cm}
Fig. \ref{fig:FeynRules} shows the Feynman diagrams for the spin-wave propagator and the two interaction vertices.
\vspace{-0.25cm}

The spin-wave coupling, $g = \frac{c}{\rho_s}$, effective Kondo coupling, $g_k = \frac{\pi}{4}(\frac{K}{c})^2 N(E_F)$, spin-wave velocity, $c = \sqrt{2d} J S a$, and spin-wave stiffness, $\rho_s = J S^2 a^{2-d}$, are derived in terms of microscopic quantities: the Heisenberg coupling $J$, Kondo coupling, $K$, lattice spacing, $a$, spin length, $S$, electronic DOS at $E_F$, $N(E_F)$, and Fermi velocity, $v_F$. For convenience, we have written $k = (\omega, \vec{k})$. 

Following methods similar to Brezin {\it et al.} \cite{BrezinPRL76,BrezinPRB76}, we define a renormalized theory at a momentum scale $\mu$, and invariance of the theory under a change of $\mu$ gives the following Callan-Symanzik (C-S) equation,
\begin{equation}
(\mu \frac{\partial}{\partial \mu} + \beta_c \frac{\partial}{\partial c} + \beta_g \frac{\partial}{\partial g} + \beta_{g_k} \frac{\partial}{\partial g_k} - \frac{N}{2} \gamma_{\pi} + \frac{N}{2} \gamma_g ) \Gamma^{(N)}_r = 0 
\label{eqn:CSeqn}
\end{equation}

\begin{figure}[hbtp!]
\begin{center}
\includegraphics[angle=0, width=5cm]{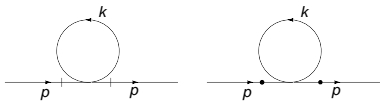}
\end{center}
\caption{The two diagrams that contribute at one-loop order to the propagator. The left diagram is the usual spin-wave correction, and the right diagram is the correction due to the Kondo interaction.}
\label{fig:propagator_1_loop_corr}
\end{figure}

We carried out a renormalization group calculation to 1-loop order within the $\epsilon$-expansion, and the system is taken to be $(1 + (1 + \epsilon))$-dimension. The 1-loop corrections to the spin-wave propagator, $G(k)$, and the Kondo coupling vertex, $\Gamma^{(4)}_{g_k}$ are calculated. The diagrams for the spin-wave propagator are shown in Fig. \ref{fig:propagator_1_loop_corr}, and there are 11 diagrams for the Kondo vertex which are not shown for convenience. We then define the following renormalization factors to absorb the divergences: the spin-wave coupling renormalization $Z_g$, spin-wave velocity renormalization $Z_c$, and the Kondo coupling renormalization $Z_{g_k}$. A straightforward perturbative calculation shows that the spin-wave velocity is renormalized, as expected from the Kondo term which breaks Lorentz-invariance. The wave-function renormalization $Z_{\pi}$ is obtained from the 1-loop correction to $\langle \sigma(x) \rangle$. We then obtain the following beta functions,
\begin{eqnarray}
\beta_g & \equiv & \frac{\partial g}{\partial \log \mu} = \epsilon g - \frac{1}{4 \pi} g^2 + \frac{1}{8 \pi} \frac{c^2}{v_F} g g _k \nonumber \\
\beta_{g_k} & \equiv & \frac{\partial g_k}{\partial \log \mu} = -\epsilon g_k + \frac{1}{4 \pi^2} \frac{v_F}{c^2} g^2 \nonumber \\
 & & \hspace{1.5cm} + \frac{4 \pi + a_1}{4 \pi^2} g g_k - \frac{a_2}{4 \pi^2} \frac{c^2}{v_F} g_k^2 \nonumber \\
\beta_c & \equiv & \frac{\partial c}{\partial \log \mu} = \frac{1}{8 \pi} \frac{c^3}{v_F} g_k \nonumber \\
\gamma_{\pi} & \equiv & \frac{\partial \log Z_{\pi}}{\partial \log \mu} = \frac{1}{2 \pi} g \nonumber \\
\gamma_{g} & \equiv & \frac{\partial \log Z_g}{\partial \log \mu} = \frac{1}{2 \pi} g - \frac{1}{8 \pi} \frac{c^2}{v_F} g_k
\label{eqn:beta-functions}
\end{eqnarray}

where $a_1 = \tfrac{1}{4 \pi^2}(2 \sqrt{\pi} + \tfrac{4}{\sqrt{\pi}} + \log(2) + \tfrac{1}{2} \psi^{(0)}(\tfrac{3}{4}) - \tfrac{1}{2} \psi^{(0)}(\tfrac{5}{4}) - 1)$ and $a_2 = \tfrac{1}{4 \pi^2}(5 \sqrt{\pi} + 2 \log(2) + \tfrac{\pi^{3/2}}{4}(2 + \psi^{(0)}(-\tfrac{1}{2}) - 2 \psi^{(0)}(\tfrac{1}{2})) - \tfrac{\pi}{4}- \tfrac{5}{2})$, and $\psi^{(0)}(z)$ is the digamma function.
\vspace{0cm}
\vspace{0cm}

\begin{figure}[hbtp!]
\begin{center}
\includegraphics[angle=-90, width=8cm]{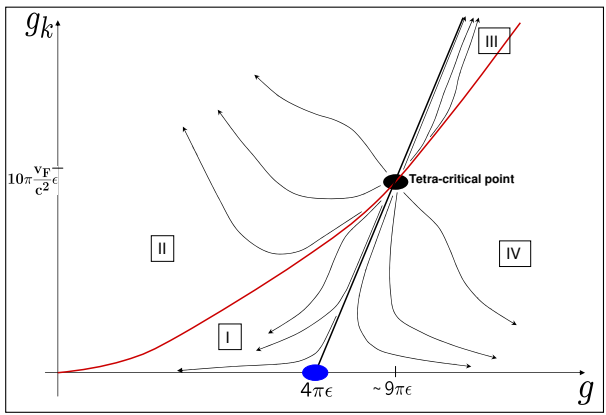}
\end{center}
\caption{Plot of RG flow.}
\label{plot:RGflow}
\end{figure}

From $\beta_g$ and $\beta_{g_k}$, we obtain the following RG-flow phase diagram, Fig. \ref{plot:RGflow}, which shows that the unstable fixed point in the NLSM has shifted to a new tetra-critical point at $g_c \approx 9.1 \pi \epsilon$ and $g_{k,c} \approx 10.2 \pi \epsilon \tfrac{v_F}{c^2}$.

Fig. \ref{plot:RGflow} shows four possible phases within reach of the $\epsilon$-expansion. In region I, both $g$ and $g_k$ flow towards zero, i.e. a stable AFM phase as expected. In region II, $g$ flows towards zero and $g_k$ flows away to a large value, indicating that the RKKY interaction between the spins is relevant. Since the RKKY coupling of the ferromagnetic moments is relevant, we expect a spiral phase with the ferromagnetic part of the spin having a wave-vector of $2 k_F$. A recent spin-polarized STM study of a lattice of manganese atoms on a tungsten surface shows a spiral phase with a wavelength of $\lambda_s$ = 12nm \cite{BodeNature07}. The Fermi wavelength is $\lambda_F$ = 20 nm, and $1/2 \lambda_F \approx \lambda_s$, indicating that it may be an experimental realization of the spiral phase of our theory (Region II).

In region IV, $g$ flows away to some large value, and $g_k$ flows to zero, which is basically identical to the quantum-disordered phase in the NLSM. Additional studies \cite{ReadPRB90, SachdevAnnPhys02} have shown that proliferation of topological excitations drive the system into a valence-bond solid state. In region III, both $g$ and $g_k$ flow away to some large value, indicating a Kondo driven paramagnetic phase where the Kondo coupling is relevant and the spins are disordered. A possibility would of course be a heavy fermi liquid PM state, but more exotic Kondo-stabilized spin liquid states as mentioned above are also possible.

We study the nature of the tetra-critical point, and show that it is an infinite-order phase transition governed by a logarithmic singularity. We solved Eq. \ref{eqn:CSeqn} and \ref{eqn:beta-functions} near the QCP, and obtained the correlation length, spin-wave velocity and propagator. The C-S equation is solved in a standard manner using the method of characteristics, and is given by
\begin{eqnarray}
\Gamma^{(2)}_r (\omega, c |\vec{k}|, g, g_k) & = & \exp^{-\int^{p}_{\mu} d \log \mu^{'} \gamma_g - \gamma_{\pi} } \nonumber \\
 & & \Gamma^{(2)}_r (\omega, c_r |\vec{k}|, g_r, g_{k,r})
\label{eqn:CSsoln}
\end{eqnarray}
where, $c_r$, $g_r$ and $g_{k,r}$ are solutions of Eq. \ref{eqn:beta-functions}, and $p = \sqrt{\omega^2 + c^ |\vec{p}|^2}$ is the momentum scale we are interested in. 

Solving the Hessian of $\beta_g$ and $\beta_{g_k}$ gives two negative eigenvalues, showing that the tetra-critical point is an unstable fixed point, and the dominant eigenvalue lies along the separatrix between regions I and IV. Since the three coupled $\beta$ functions are difficult to solve in general near the critical point, we will solve it along the direction of the dominant eigenvalue, where we know $g$ as a function of $g_k$ and $c$. Linearizing the $\beta$ functions we are able to obtain the following solutions.
\begin{eqnarray}
c(\mu) & \approx & \frac{c_0}{\sqrt{1 + \frac{1}{2} \log \frac{y(\mu)}{y_0}}} \nonumber \\
y(\mu) & \approx & y_0 \left( 2 ProductLog \left( \frac{1}{2} \exp \left( \frac{1}{2}(\frac{\Lambda}{\mu})^{\frac{5}{2 \epsilon}} \right) \right) \right)^2
\label{eqn:ysoln}
\end{eqnarray}

$\Lambda$ is the cutoff of the bare theory with the unrenormalized parameters. Here $y$ is defined as $g_k = g_{k,c}(1-y)$, and $y_0$ measures the initial distance from the critical point. The function $ProductLog(x)$ is also known as the Lambert function, $W(x)$. From Eq. \ref{eqn:ysoln}, we obtain the correlation length, $\xi$ when $y(\mu) \sim 1$, and $a_0 \sim \Lambda^{-1}$.
\begin{equation}
\frac{\xi}{a_0} \approx \left( \frac{y(\mu)}{y_0} \right)^{\frac{1}{5 \epsilon}} \left( \log \frac{y(\mu)}{y_0} \right)^{\frac{2}{5\epsilon}}
\label{eqn:corrlength}
\end{equation}

This shows that we have an infinite-order phase transition that is dominated by a logarithmic singularity. The phase transition due to spin-wave fluctuations seen in the NLSM is now dominated instead by Kondo fluctuations of the spins. This clearly shows that local Kondo fluctuations play an essential role in driving the QPT, and is reflected in the Kondo interaction being relevant for the two phases found in region II and III.

We also solved for the spin-wave propagator near the QCP using Eq. \ref{eqn:CSsoln}, which gives,
\begin{equation}
\Gamma^{(2)}_r \approx \left( \frac{y(\mu)}{y_0} \left( \log \frac{y(\mu)}{y_0} \right)^{2} \right)^{\frac{1}{2} + \frac{1}{5}}\left( \omega^2 + \frac{c_0^2 |\vec{k}|^2}{1+\frac{1}{2} \log \frac{y(\mu)}{y_0}} \right)
\label{eqn:RenProp}
\end{equation}

Note that in Eqns. \ref{eqn:ysoln} - \ref{eqn:RenProp}, the exponents are numerical values that depend on $a_1$ and $a_2$; the actual values are $\tfrac{1}{2} \approx \tfrac{1}{1.96}$ and $\tfrac{1}{5} \approx \tfrac{1}{4.97}$, which we closely round off in the exponents.

Eq. \ref{eqn:ysoln} shows that for small enough momentum, $p \ll (e + 2)^{-\frac{2}{5 \epsilon}} \Lambda$, the spin wave velocity scales logarithmically, which can be seen as a cross-over length-scale from Gaussian behavior to critical behavior of the spin waves. Similar behavior occurs in the Grinstein-Pelcovits renormalization of the elastic constants in smectic crystals \cite{GrinsteinPRA82}. Therefore for a small enough region close to the QCP and small enough momentum, the spin-wave velocity scales logarithmically as,
\begin{equation}
c(\mu) \sim \frac{c_0}{\sqrt{\log \frac{y(\mu)}{y_0}}}
\label{eqn:cAsy}
\end{equation}

Thus for asymptotically small momentum, the spin-wave velocity vanishes logarithmically, indicating breakdown of hydrodynamic behavior. This implies that the spin waves become increasingly localized near the QCP, and that localized Kondo-induced spin flips that fluctuate temporally may also be critical modes at the QCP, which may be a mechanism for driving the system into a heavy fermion or new exotic phase. Similar ideas of local quantum criticality have been proposed by other groups \cite{SiNature01, GrempelPRL03}, and a similar study using Shankar's fermionic RG method found a Lifshitz transition \cite{YamamotoPRL07}.

In conclusion, we have studied the antiferromagnetic quantum phase transition of a Kondo lattice with a small Fermi surface. Since the Fermi surface has no ``hot"-spots nested by $(\pi,\pi)$, there is no Landau damping of the spin waves, and the theory remains $z = 1$. The effective theory contains a new Kondo driven interaction that significantly renormalizes the spin wave velocity, and is essential in driving the quantum phase transition. 

The RG results suggest a possible phase diagram with two disordered phases, one with the Kondo coupling relevant, and the other with the Kondo coupling irrelevant, where exotic phases with fractionalized quasi-particles may be found. One result of our calculations is that the Kondo coupling is irrelevant in the magnetic phase, but is relevant in one of the disordered phases, which is in agreement with some of the experimental results showing a change in Fermi surface size \cite{PaschenNature04}. 

The key result of our RG calculations is that the QCP of the NLSM is now shifted to a Kondo-driven infinite-order QCP that is characterized by a logarithmic singularity. The velocity of the critical modes are logarithmically slowed down due to Kondo fluctuations with the conduction electrons near the QCP, leading to localization at sufficiently small momenta. This clearly shows that local Kondo physics plays an essential role in the quantum phase transition, as opposed to the standard view of a spin wave driven QPT that is seen in the NLSM. The logarithmic scaling of the spin-wave velocity is a quantum analog of a similar logarithmic scaling of the elastic coefficient seen in a classical Lifshitz transition. The appearance of localized critical modes at a QCP, termed local quantum criticality, is a topic of great current interest. An outstanding issue is the effects of topological excitations on the QCP that we have found, especially in the presence of Kondo coupling to conduction electrons. Our work could be of relevance to heavy fermion systems that display effective 2D-like transitions, and also to STM-engineered 2D Kondo lattices.

\begin{acknowledgements}
We would like to thank Steve Kivelson, Steve Shenker and Eduardo Fradkin for many enlightening discussions that were essential for this work. This work was partly supported by the IBM Almaden-Stanford Student award.
\end{acknowledgements}

\bibliographystyle{btxbst}

\end{document}